\documentclass{osa-article}

\journal{boe}

\articletype{Research Article}

\begin{document}

\title{High signal-to-noise ratio reconstruction of low bit-depth optical coherence tomography using deep learning}

\author{Qiangjiang Hao,\authormark{1,5,6} Kang Zhou,\authormark{1,2,6} Jianlong Yang,\authormark{1,*} Yan Hu,\authormark{3} Liyang Fang,\authormark{1} Zhengjie Chai,\authormark{2} Yuhui Ma,\authormark{1} Gangjun Liu,\authormark{4} Shenghua Gao,\authormark{2} and Jiang Liu\authormark{1,3}}
\address{\authormark{1}Cixi Institute of Biomedical Engineering, Ningbo Institute of Materials Technology and Engineering, Chinese Academy of Sciences, China\\
\authormark{2}School of Information Science and Technology, ShanghaiTech University, China\\
\authormark{3}Department of Computer Science and Engineering, Southern University of Science and Technology, China\\
\authormark{4}Shenzhen Bay laboratory, China\\
\authormark{5}Nano Science and Technology Institute of
USTC, China\\
\authormark{6} Equally contribution\\
\email{\authormark{*} yangjianlong@nimte.ac.cn}}



\begin{abstract}
Reducing the bit-depth is an effective approach to lower the cost of optical coherence tomography (OCT) systems and increase the transmission efficiency in data acquisition and telemedicine. However, a low bit-depth will lead to the degeneration of the detection sensitivity thus reduce the signal-to-noise ratio (SNR) of OCT images. In this paper, we propose to use deep learning for the reconstruction of the high SNR OCT images from the low bit-depth acquisition. Its feasibility was preliminarily evaluated by applying the proposed method to the quantized $3\sim8$-bit data from native 12-bit interference fringes. We employed a pixel-to-pixel generative adversarial network architecture in the low to high bit-depth OCT image transition. Retinal OCT data of a healthy subject from a homemade spectral-domain OCT system was used in the study. Extensively qualitative and quantitative results show this deep-learning-based approach could significantly improve the SNR of the low bit-depth OCT images especially at the choroidal region. Superior similarity and SNR between the reconstructed images and the original 12-bit OCT images could be derived when the bit-depth $\geq 5$. This work demonstrates the proper integration of OCT and deep learning could benefit the development of healthcare in low-resource settings.
\end{abstract}

\section{Introduction}
Optical coherence tomography (OCT) is a non-invasive cross-sectional high-resolution imaging modality that has been widely used in various medical fields, such as ophthalmology, cardiovascular endoscopy, and dermatology \cite{Swanson2017}. In ophthalmology, OCT has become the clinical standard for the examination of non-superficial retinal lesions, such as choroidal neovascularization, macular edema, and pigment epithelial detachment \cite{adhi2013optical}. In the technical aspect, the acquisition speed of OCT systems keeps increasing from tens of hertz at the beginning of its invention to several megahertz nowadays \cite{Klein2017}, which enables the OCT imaging to have fewer motion artifacts, wider field of view, better resolutions, and higher detection sensitivity. \\
\indent However, as the OCT systems become faster, the data acquisition and transmission of the OCT scans become problematic. For example, OCT acquisition usually has the bit-depth of 12 or 14 bit, so 1-gigabyte data will be generated during a typical volumetric scan with 2048 data points in axial directions and 512 data points in both lateral directions. This large amount of data requires the analog-to-digital converter (ADC) to have high digitizing speed and large on-board memory, which will significantly increase the cost of the OCT systems thus hinder their popularization in clinical applications. On the other hand, the large sizes of the OCT data bring difficulties to their transmission, so the telemedicine that bridges the regions in shortage of medical resources and the medical experts is hard to implement.\\
\indent To reduce the sizes of the OCT data, researchers have tried to first decrease the spatial sampling rate below the Nyquist-Shannon limit then reconstruct the data using compressed sensing (CS) techniques \cite{Graff2015}. Liu and Kang applied pseudo-random masks to sample part of the CCD pixels then reconstruct the $k$-space signal by minimizing the $l_1$ norm of a transformed image to enforce sparsity, subject to data consistency constraints \cite{Liu2010}. Lebed \textit{et al.} proposed a volumetric scan pattern that composed randomly spaced horizontal and vertical B-scans for the CS reconstruction \cite{Lebed2010} then they used this method in the real-time 3D imaging of the optic nerve head by a spectral domain (SD) OCT system \cite{Young2011}. Zhang \textit{et al.} employed a $k$-linear mask to sample the OCT interferogram evenly spaced in the wavenumber domain, which could use less than 20\% of the total data and get rid of the spectral calibration and interpolation processes \cite{Zhang2012}. Fang \textit{et al.} reconstructed the low transverse sampled OCT images using sparse representation \cite{Fang2013}. \\
\begin{figure}[h!]
\centering\includegraphics[width=12cm]{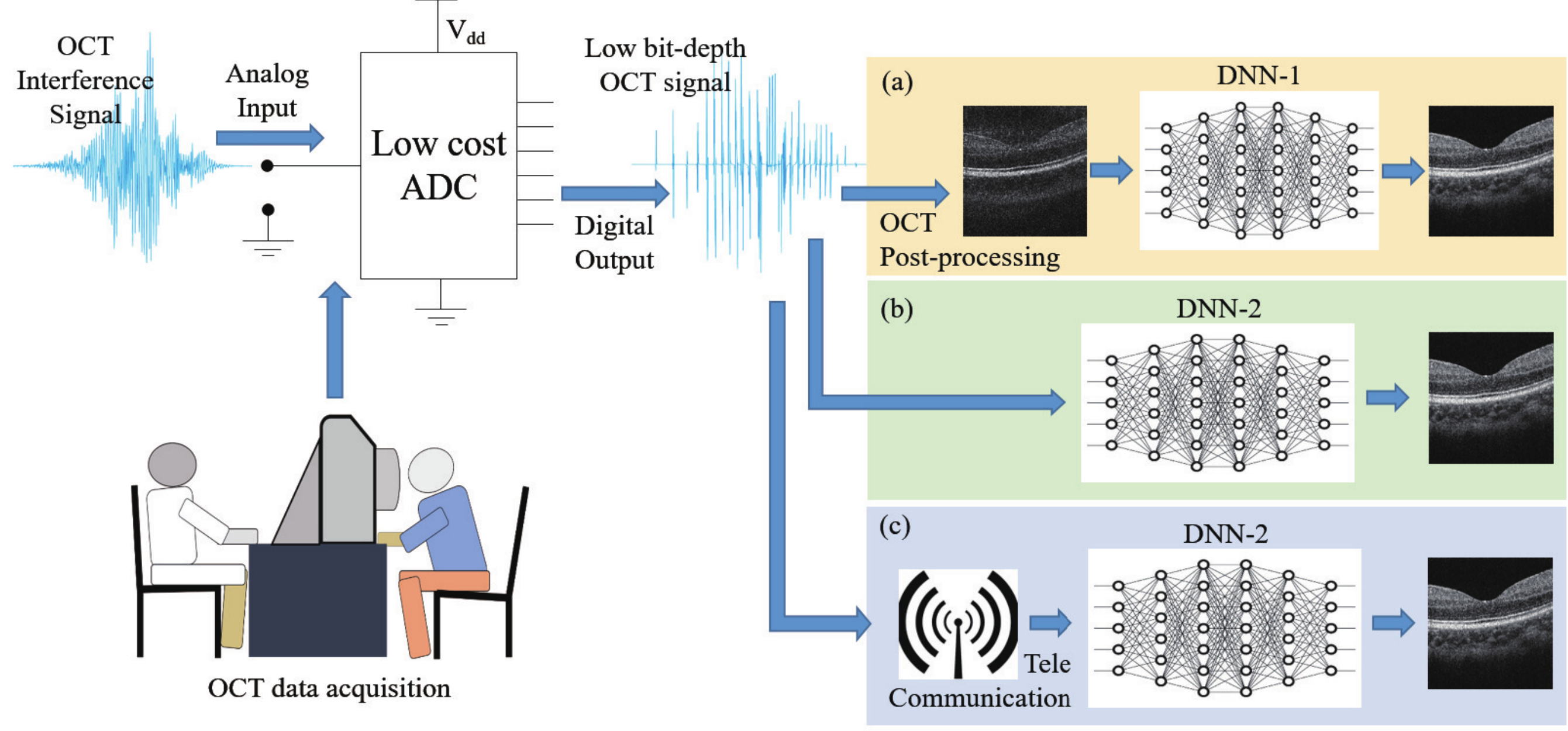}
\caption{Schematic of reconstructing high SNR OCT images from low bit-depth signals using deep learning. (a) Converting the low bit-depth signals into OCT images then using a deep neural network (DNN-1) to generate high SNR OCT images. (b) Directly converting the low bit-depth signals into high SNR images using the DNN-2. (c) Using telecommunication to transmit the low bit-depth interferograms to the servers of medical experts then converting them into OCT images using the DNN-2.}
\label{fig1}
\end{figure}
\indent Although the spatial undersampling for the OCT compression has been well explored, there is no attempt to reconstruct the OCT images from a low bit-depth data (undersampling in intensity), to the best of our knowledge. Even though the influences of the bit-depth on OCT imaging have been extensively investigated by several groups \cite{Goldberg2009,Lu2011,Ling2012}. Goldberg \textit{et al.} used a swept source (SS) OCT system for human coronary imaging and studied the signal-to-noise ratio (SNR), sensitivity, and dynamic range as a function of the bit-depth \cite{Goldberg2009}. They found the SNR increased as the bit-depth increased but trended to be stable when the bit-depth $\geq 8$. Lu \textit{et al.} compared the performance of an 8-bit ADC and a 14-bit ADC in a polarization sensitive SS OCT system and found the sensitivity and dynamic range drops due to the low bit-depth \cite{Lu2011}. Ling and Ellerbee studied the effects of the low bit-depth on the phase of the OCT data and demonstrated the phase noise could be significantly magnified as the bit-depth decreased \cite{Ling2012}.\\
\indent Here we propose to compress the OCT data by reducing the acquisition bit-depth. We further propose to employ the emerging deep learning techniques to compensate for the data quality degeneration caused by the low bit-depth mentioned above. Deep learning techniques have been successfully used in the data reconstruction of medical imaging modalities like magnetic resonance imaging and low-dose X-ray computed tomography \cite{Zhu2018,Han2019}. They also have employed in other types of OCT reconstruction including denoising \cite{ma2018speckle}  and super-resolution \cite{huang2019simultaneous}.  However, we did not find the works related to the low bit-depth reconstruction of OCT data in the literature.\\
\indent The ultimate goal of this study is using the high SNR reconstruction of the low bit-depth OCT to benefit the popularization of this technique (reduce its cost) and the telemedicine as shown in Fig.~\ref{fig1}(c). The previous step of this goal is to convert the original low bit-depth interference signals to the high SNR OCT images through a deep neural network (DNN) as shown in Fig.~\ref{fig1}(b). Figure~\ref{fig1}(a) gives an alternative approach for OCT data compression in telemedicine suggested by Mousavi \textit{et al.} \cite{Mousavi2014}, which converts the interference signals into OCT images before feeds them into the DNN. Serial number 1 and 2 are used to differentiate the DNN used in the image to image conversion from the DNN used in the interferogram to image conversion.\\
\indent In this paper, we preliminarily evaluated the feasibility of the proposed idea by converting the low bit-depth OCT images to high bit-depth OCT images using a generative adversarial network (GAN). The original 12-bit OCT interferograms was requantized into $3\sim8$-bit fringes and converted to the OCT images with different bit-depths using standard OCT postprocessing method. We qualitatively and quantitatively compared the deep-learning-converted OCT images with the native 12-bit OCT images. The results show the proposed approach is promising in the reconstruction of high SNR OCT from low bit-depth images.
\section{Materials and methods}
\subsection{Data preparation}
We employed a homemade spectral domain OCT system with a A-line rate of 70 kHz. A 24 years old healthy volunteer was recruited in this study. The acquired OCT volumes have a field-of-view of $3\times3$ mm$^2$ and were centered on the fovea. An equivalent sampling of 200 was used along the two transverse directions. Each A-line has 1024 pixels. The human study protocol was approved by the Institutional Review Board of Cixi Institute of Biomedical Engineering, Chinese Academy of Sciences and followed the tenets of the Declaration of Helsinki. \\
\indent To achieve the lower bit-depth images, we quantized the raw 12-bit interference fringes to simulate the sampling depths ranging from 3 to 8 bit with an increment of 1 bit \cite{ling2012effects}.\\
\begin{figure}[h!]
\centering\includegraphics[width=11.9cm]{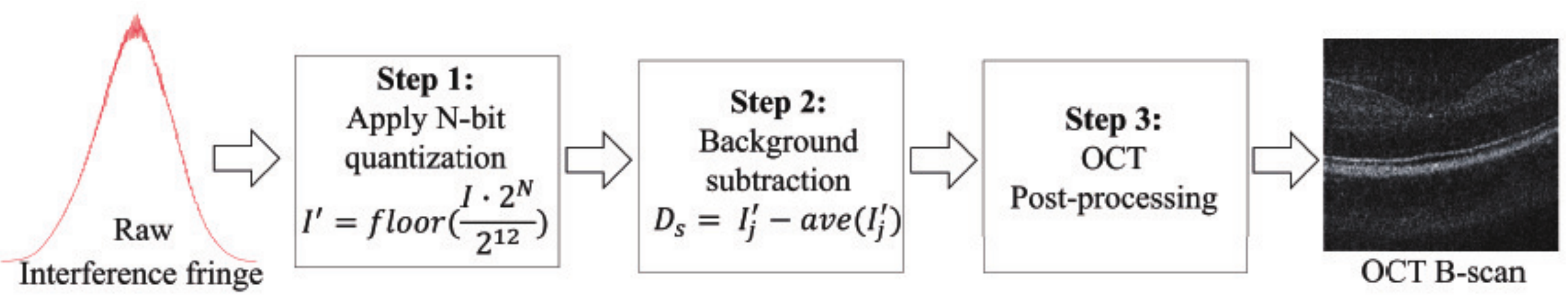}
\caption{Method to generate the low bit-depth OCT images.}
\end{figure}\label{simu}
 \begin{figure}[t!]
\centering\includegraphics[width=12cm]{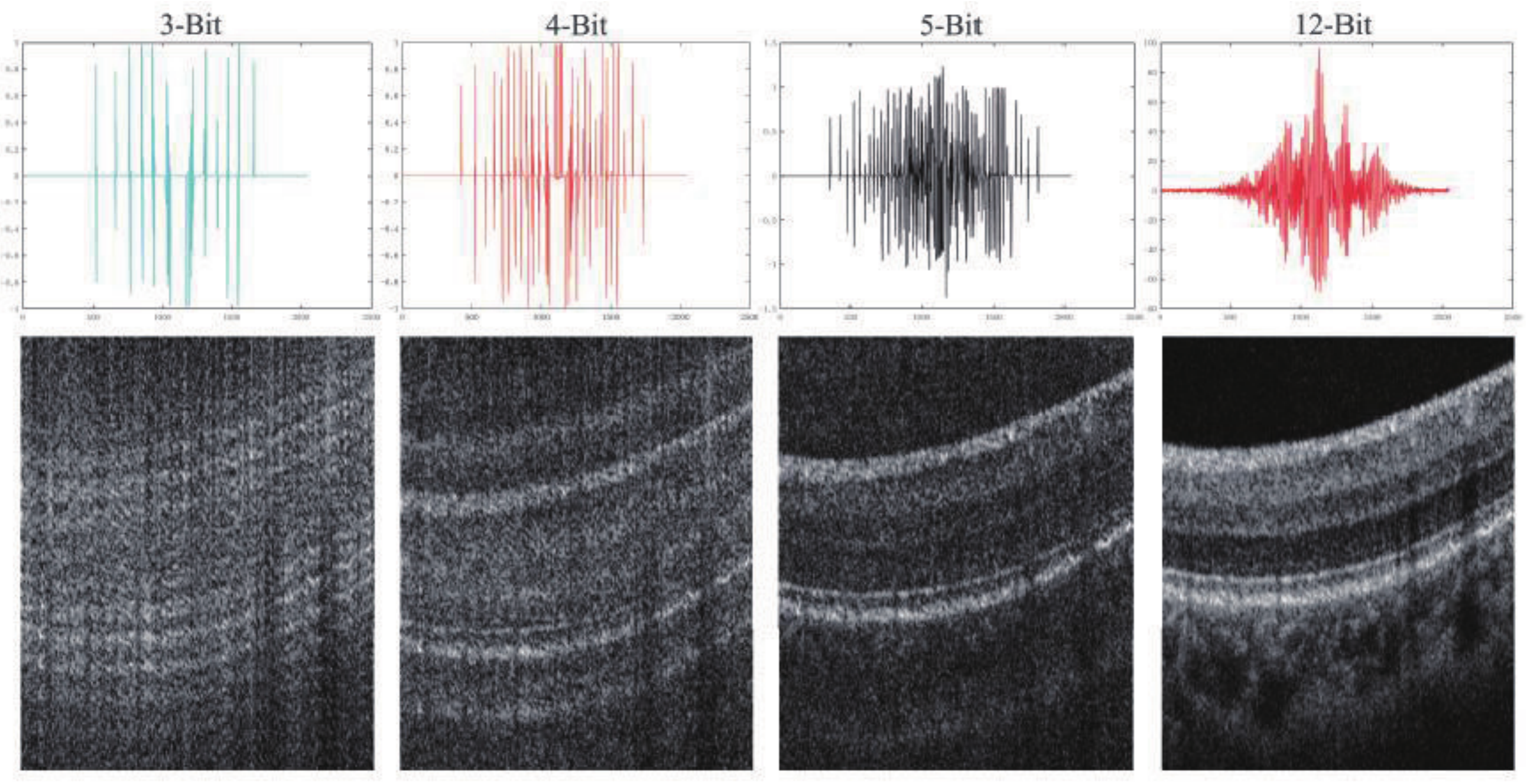}
\caption{Different bit-depth digital signals correspond to different quality OCT B-scan images.}\label{fig3}
\end{figure}
\indent Figure~\ref{simu} illustrates the steps involved. The original data from the ADC board was read out as the integral values ranging from 0 to $2^{12}-1$.  For each bit-depth level, the intensity values of the interference signal were convert from the original 12-bit values using \cite{goldberg2009performance}:
\begin{equation}
I^{\prime}=floor \operatorname{}\left(\frac{I \cdot 2^{N}}{2^{12}}\right)
\end{equation}
Where $floor$ is the floor function that rounds toward negative infinity. $N$ represents the different bit depth, $I$ presents the raw data, while $I^{\prime}$ presents the converted data. Then the background is removed by
\begin{equation}
D_{S}=I_{j}^{\prime}-ave\left(I_{j}^{\prime}\right)
\end{equation}
Where $D_{S}$ presents the signal to remove the background, $j$ presents every column of the data $I^{\prime}$, function $ave$ calculates the average of each column of the raw matrix. Then the converted signal was processed using the OCT post-processing pipeline including background subtraction, $k$-linearization, dispersion compensation, Fourier transformation, and image logarithm. Figure~\ref{fig3} demonstrates different bit-depth digital signals correspond to different quality OCT B-scan images.
\subsection{Deep learning network}
\begin{figure}[h!]
\centering\includegraphics[width=12cm]{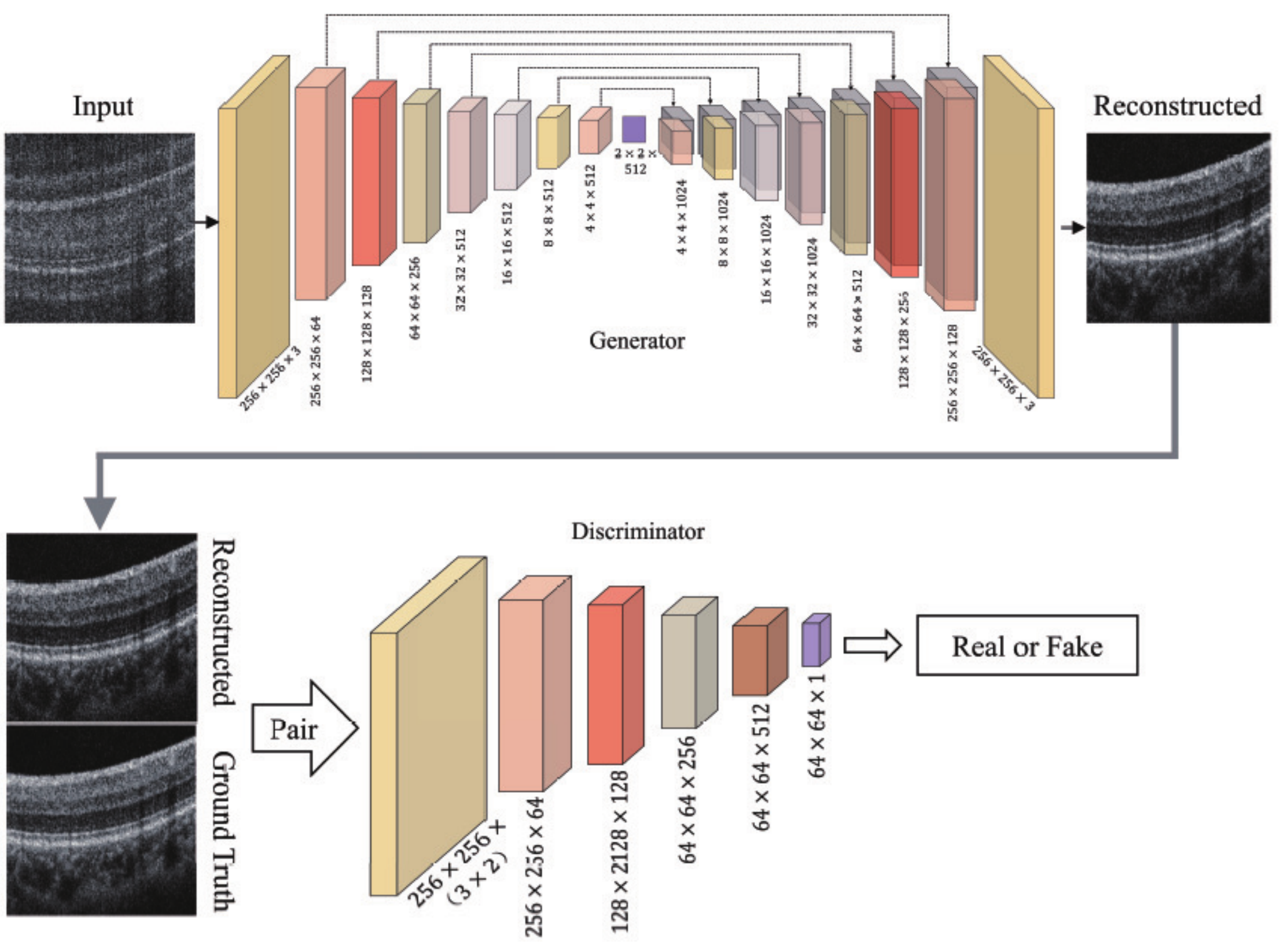}
\caption{Illustration of the proposed framework for the high SNR reconstruction of the low bit-depth OCT images.}\label{pix}
\end{figure}

\indent In this paper, we propose to use the pixel-to-pixel GAN architecture (pix2pixGAN) \cite{isola2017image} for the low to high bit-depth transition, because the structure and texture information of the low bit-depth images and the high bit-depth images are precisely corresponding.\\
\indent The overall framework is illustrated in Fig~\ref{pix}. In this framework, the generator is implemented by the U-shape network architecture, which can prevent losing small objects because of the skip connection between each centrally symmetric layers \cite{sato2018segmentation}. As for the discriminator, we adopt the PatchGAN \cite{isola2017image}, which models the OCT image as a Markov random field and only penalizes structure at the scale of patches. The patchGAN  restricts the attention of the discriminator to high-frequency, so it can avoid blurry results.  Also, using the patches instead of the entire image can reduce the number of parameters and accelerate the training.
\\
\indent The objective of the pix2pixGAN is expressed as\cite{isola2017image}:
\begin{equation}
\mathcal{L}_{ G A N}(G, D)=E_{x, y}[\log D(x, y)]+E_{x}[\log (1-D(x, G(x)))],
\end{equation}
where the generator $G$ tries to minimize this objective against an adversarial discriminator $D$ tries to maximize it, $x$ is the input low-bit OCT B-scan image, $y$ is the corresponding 12-bit depth OCT B-scan set as the ground truth for $x$. During the training process\cite{ma2018speckle}, $G$ tries to minimize the goal, while $D$ tries to maximize the goal, so the results are optimized:
\begin{equation}
G^{*}=\arg \min _{G} \max _{D} \mathcal{L}_{ G A N}(G, D),
\end{equation}
where $G^{*}$ means the resulted optimized generator.\\
\indent The purpose of the discriminator remains the same, but the task of the generator is not only to trick the discriminator, but also to approach $L2$'s ground real-time output so we use $L1$ distance instead of $L2$: 
\begin{equation}
\mathcal{L}_{L 1}(G)=E_{x, y}\left[\|y-G(x)\|_{1}\right].
\end{equation}
So the final objective is:
\begin{equation}
G^{*}=\arg \min _{G} \max _{D} \mathcal{L}_{ G A N}(G, D)+\lambda\mathcal{L}_{L 1} (G)
\end{equation}
\subsection{Implementation}
The code was implenented in PyTorch and trained on a personal workstation using the NVIDIA GTX 1080ti GPU with 12 GB memory and the operating system is Ubuntu 16.04 LTS. To optimize the deep neural network, we adopt the standard approaches from \cite{isola2017image}. We set the epoch $= 200$ and batch size $= 1$. The hyperparameter $\lambda = 10$. We used the Adam solver \cite{kingma2014adam} to train generator from scratch. The initial learning rate was set as $2\times10^{-4}$. For training the discriminator, we also adopt the Adam optimizer \cite{kingma2014adam} with the learning rate as $2\times10^{-4}$. We set $\beta_1=0.5$, $\beta_2=0.999$ for both of the two Adam optimizers.\\
\indent In our experiment, we achieved 200 B-scans at each bit depth using the method described in Section 2.1. These images were divided into 160 training images, 20 validation images, and 20 test images randomly. The size of the original B-scans is $200\times1024$ pixels, we resized them to $256\times256$ pixels for the convenience of the network training. We trained the model from scratch without data argumentation. The model converges at around 100 epochs and costs around 5 hours for each training.
\subsection{Quantitative evaluation metrics}
We employed three metrics in the quantitative comparison: peak signal-to-nise ratio (PSNR), multi-scale structural similarity index (MSSSIM), and 2D correlation coefficient (CORR2) \cite{hore2010image}. \\
\indent The PSNR is an objective standard for evaluating the SNR of an image. It is the ratio between the maximum signal and background noise. The values of the PSNR are in direct proportion to the SNR of an image. It is defined as
\begin{equation}
\mathrm{PSNR}=10 \cdot \log _{10}\left(\frac{M A X_{I}^{2}}{M S E}\right),
\end{equation}
where the $M A X_{I}$ is the maximum value of the intensity in the OCT images, and MSE is the mean squared error.
\\
\indent
We employ the MSSSIM to indicate the similarity between two images. Compared with the single-scale structural similarity index, the MSSSIM supplies more flexibility in incorporating the variations of viewing conditions\cite{wang2003multiscale}. The measures of luminance $L$, contrast $C$ and structure comparison $S$ are defined as follows:
\begin{equation}
\mathrm{L}(\mathrm{X}, \mathrm{Y})=\frac{2 u_{X} u_{Y}+C_{1}}{u_{X}^{2}+u_{Y}^{2}+C_{1}},
\end{equation}
\begin{equation}
C(X, Y)=\frac{2 \sigma_{X} \sigma_{Y}+C_{2}}{\sigma_{X}^{2}+\sigma_{Y}^{2}+C_{2}},
\end{equation}
\begin{equation}
\mathrm{S}(\mathrm{X}, \mathrm{Y})=\frac{\sigma_{X Y}+C_{3}}{\sigma_{X} \sigma_{Y}+C_{3}},
\end{equation}
where $X$ is the tested image and $Y$ is the reference image. $u_{X}$ and $u_{Y}$ are their mean values. $\sigma_{X}$ and $\sigma_{Y}$ are their standard deviations. $C_{1}$, $C_{2}$ and $C_{3}$ are small constants, we take $C_{1}=10^{-4}$, $C_{2}=10^{-4}$ and $C_{3}=0.5C_{2}$ here.\\
\indent So the overall MSSSIM evaluation is the combination of these measures at different scales:
\begin{equation}
\operatorname{MSSSIM}(\mathrm{X}, \mathrm{Y})=\left[L_{M}(X, Y)\right]^{\alpha_M} \prod_{j=1}^{M}\left[C_{j}(X, Y)\right]^{\beta_{j}}\left[S_{J}(X, Y)\right]^{\gamma_{j}}.
\end{equation}
The exponents $\alpha_M$, $\beta_{j}$, and $\gamma_{j}$ are used to adjust the relative importance of different components. We take $M=1$, ${\alpha=1}$, ${\beta=\gamma=0.0448}$. \\
\indent The CORR2 function implements the Pearson correlation to 2D arrays\cite{ramadan2017using} between images A and B. The function is defined as:
\begin{equation}
\mathrm{CORR2}=\frac{\sum_{m} \sum_{n}\left(A_{m n}-\overline{A}\right)\left(B_{m n}-\overline{B}\right)}{\sqrt{\left(\sum_{m} \sum_{n}\left(A_{m n}-\overline{A}\right)^{2}\right)\left(\sum_{m} \sum_{n}\left(B_{m n}-\overline{B}\right)^{2}\right)}},
\end{equation}
where $A_{mn}$ is the intensity of the $(m,n)$ pixel in the image A, $B_{mn}$ is the intensity of the $(m, n)$ the pixel in the image B, $\overline{A}$ is the average intensity of the image A, and $\overline{B}$ is the average intensity of the image B.
\section{Results}
\subsection{Qualitative evaluation}
\begin{figure}[h!]
\centering\includegraphics[width=10cm]{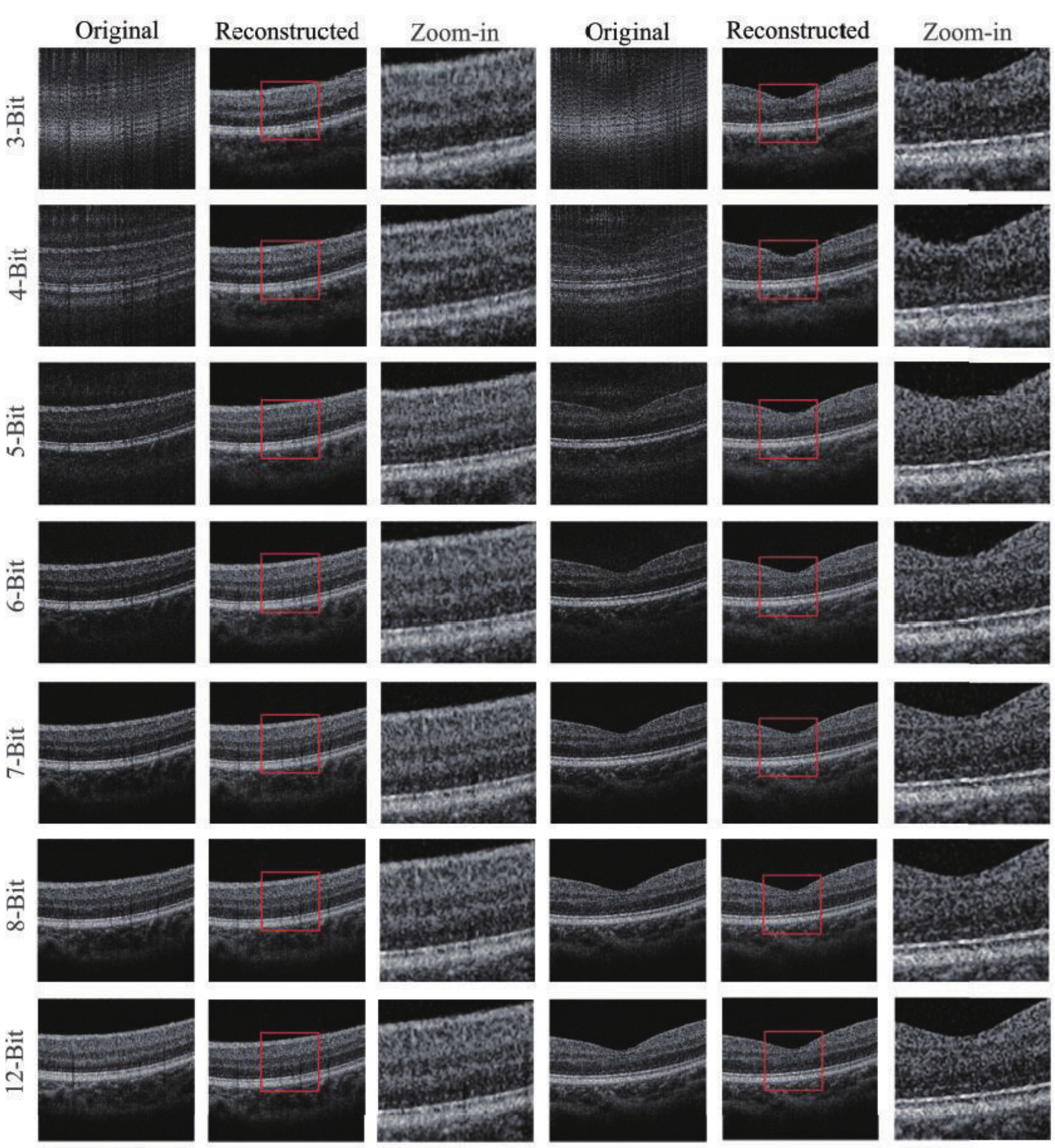}
\caption{Visual comparison of the original OCT images with different bit depths, their corresponding GAN-reconstructed images, and the zoom-in views of the reconstructed images in the red boxes.}\label{fig5}
\end{figure}
Figure~\ref{fig5} demonstrates the visual comparison of the original OCT images with different bit depths, their corresponding GAN-reconstructed images, and the zoom-in views of the reconstructed images in the red boxes. The Original 3-bit and 4-bit images are quite blurry and are unable to visualize the retinal layers clearly. When the bit-depth increases to 5, the blur of the image disappears but the SNR is very low except the high reflective inner limiting membrane (ILM) and retinal pigment epithelium (RPE) layers. As the bit-depth further increases to $6\sim 7$, the  SNR of the image keeps improving especially the visibility of the choroid. When the bit-depth $\geq 8$, further increase of the bit-depth only slightly improve the SNR. The 8-bit OCT B-scans have good similarity compared with the 12-bit images.\\
\indent The reconstructed images, on the other hand, are presented without the blur and low SNR of the original images even at the 3-bit sampling. All of the reconstructed images have good visibility of the retinal layers and the choroid and excellent similarity compared with the 12-bit images. We further compared the zoom-in view of the red boxes inside the reconstructed images. When the images with 3 and 4-bit depths, the inaccurate reconstruction exists at the outer plexiform and nuclear layers of the left side image and the foveal position of the right side image. When the bit depth $\geq 5$, the reconstructed images are quite close to the 12-bit OCT B-scans.
\subsection{Quantitative evaluation}
\begin{table}[h!]
\centering \caption{Comparison of the quantitaitve metrics from the original and reconstructed images with different bit depths.}
\begin{tabular}{ccccc}
\hline
{\bf Bit depth} & {\bf image} & {\bf PSNR (dB)} & {\bf MSSSIM} & {\bf CORR2}\\
\hline
\multicolumn{ 1}{c}{3-bit} &   Original & 16.494$\pm$0.492 & 0.573$\pm$0.010 & 0.638$\pm$0.025  \\

\multicolumn{ 1}{c}{} & Reconstructed & 20.495$\pm$0.270 & 0.786$\pm$0.015 & 0.880$\pm$0.021  \\
\hline
\multicolumn{ 1}{c}{4-bit} &   Original & 17.276$\pm$0.473 & 0.625$\pm$0.027 & 0.675$\pm$0.026  \\

\multicolumn{ 1}{c}{} & Reconstructed & 21.387$\pm$0.356 & 0.814$\pm$0.013 & 0.905$\pm$0.015  \\
\hline
\multicolumn{ 1}{c}{5-bit} &   Original & 19.499$\pm$0.339 & 0.787$\pm$0.020 & 0.926$\pm$0.026  \\

\multicolumn{ 1}{c}{} & Reconstructed & 23.389$\pm$0.386 & 0.866$\pm$0.010 & 0.951$\pm$0.010  \\
\hline
\multicolumn{ 1}{c}{6-bit} &   Original & 23.309$\pm$0.344 & 0.849$\pm$0.016 & 0.938$\pm$0.009  \\

\multicolumn{ 1}{c}{} & Reconstructed & 26.239$\pm$0.402 & 0.871$\pm$0.010 & 0.969$\pm$0.005  \\
\hline
\multicolumn{ 1}{c}{7-bit} &   Original & 27.106$\pm$0.401 & 0.944$\pm$0.006 & 0.983$\pm$0.002  \\

\multicolumn{ 1}{c}{} & Reconstructed & 29.402$\pm$0.679 & 0.944$\pm$0.007 & 0.985$\pm$0.003 \\
\hline
\multicolumn{ 1}{c}{8-bit} &   Original & 33.788$\pm$0.517 & 0.982$\pm$0.002 & 0.995$\pm$0.001  \\

\multicolumn{ 1}{c}{} & Reconstructed & 34.365$\pm$0.568 & 0.975$\pm$0.004 & 0.994$\pm$0.002  \\
\hline
\end{tabular}\label{table1} 
\end{table}
\begin{figure}[h!]
\centering\includegraphics[width=11cm]{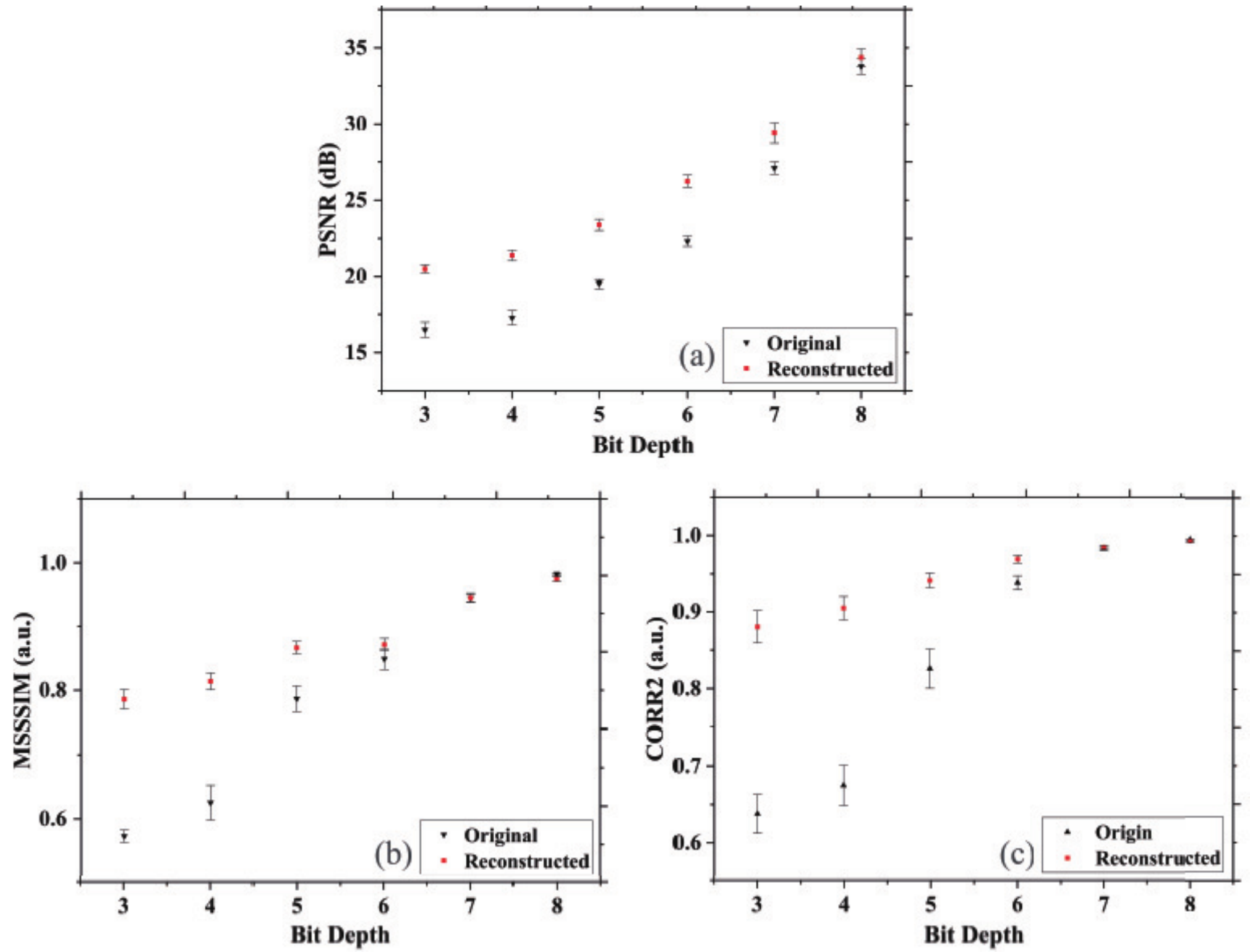}
\caption{The PSNR, MSSSIM and CORR2 as functions of bit depth. The black triangles represents the values of the original images. The red boxes represents the values of the reconstructed images.}\label{fig6}
\end{figure}
\indent We then evaluated this deep-learning-based reconstruction using the quantitative metrics defined in Section 2.4. We used the 12-bit images as the reference in the calculation. The PSNR is capable of characterizing the enhancement of the SNR. As shown in Table~\ref{table1}, We can see the reconstruction can significantly raise the PSNR when the bit depths are low. As the bit-depth increases to 8, the improvement of the reconstruction becomes marginal, because the original images are very similar to the reference. The MSSSIM and CORR2 are the metrics of similarity from different aspects. As the bit-depth increases, these metrics keep raising and getting close to 1. 
The deep-learning-based reconstruction can significantly improve the similarity of the low-bit depth images. The original high bit-depth images have good similarity thus the room for improvement becomes low. Even so, the proposed method still can enhance their similarity with the 12-bit reference. When the bit-depth $\geq 5$, the MSSSIM $>0.85$ and the CORR2 $>0.95$, which indicates the accurate reconstruction of the OCT images.\\
\indent Figure~\ref{fig6} demonstrates the calculated quantitative metrics as the functions of bit-depth. As the bit-depth increases, for each metric, the difference of the original and reconstructed images keep decreasing and converge at the high bit-depths. For the two similarity metrics, the MSSSIM and CORR2, there is a leap from the values of the bit-depth of $3\sim4$ to that of the bit-depth of 5, which corresponds to the conversion of the blur OCT B-scans to clear low SNR images as shown in Fig.~\ref{fig5}. The metric values of the reconstructed images are always higher than those of the original images, which demonstrates the effectiveness of the proposed method.
\subsection{Comparison with other deep learning methods}
\begin{figure}[h!]
\centering\includegraphics[width=11cm]{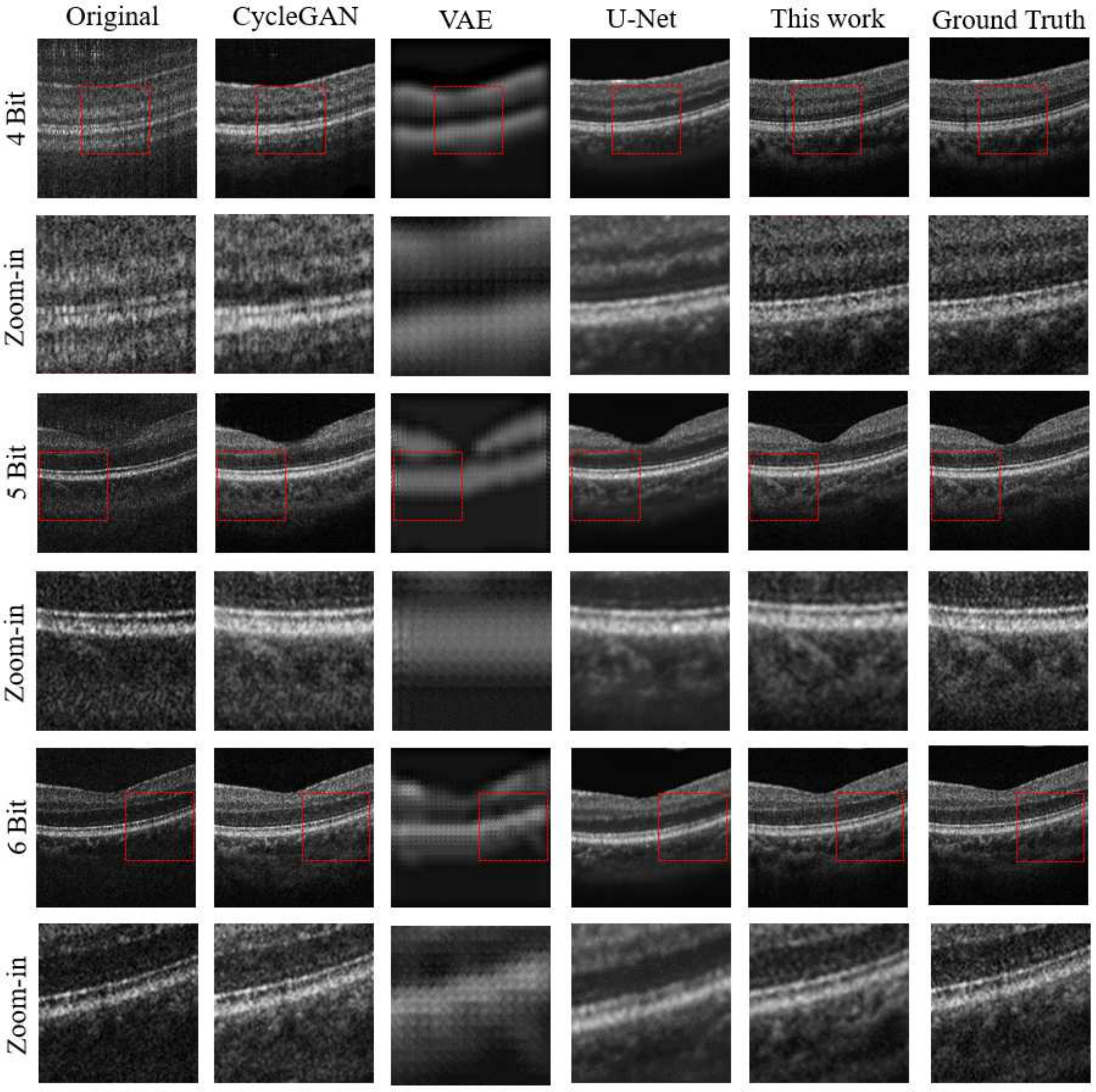}
\caption{Visual examples of the low-bit-depth OCT reconstruction using different deep learning methods. From left to right: original images, the reconstruction results using the cycleGAN, VAE, U-Net, and the pix2pixGAN adopted in this work, and the ground truth. Row 1, 3, and 5 are the results using 4-bit, 5-bit, and 6-bit images, respectively, row 2, 4, and 6 are their corresponding zoom-in views inside the red boxes.}\label{comDeep}
\end{figure}
\begin{figure}[h!]
\centering\includegraphics[width=11cm]{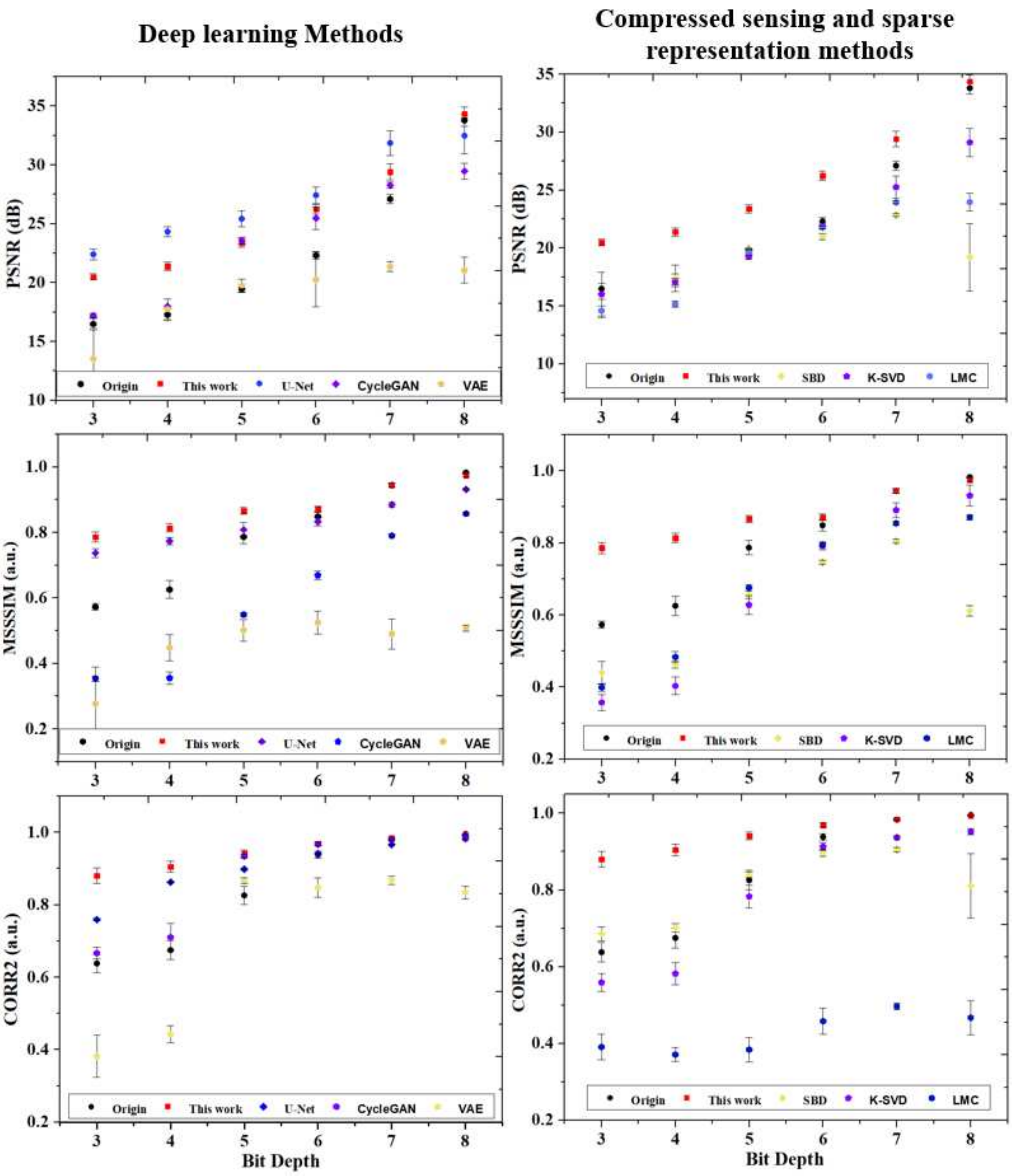}
\caption{Quantitative metrics of the low-bit-depth OCT reconstruction using different methods. From top to bottom are the PSNR, MSSSIM, and CORR2 as functions of the bit depth. The left column are the results using deep learning methods. The black dots, red blocks, blue diamonds, purple pentagons, and yellow pentagons are the results of the original images, this work, U-Net, cycleGAN, and VAE, respectively. The right column are the results using compressed sensing and sparse representation methods. The black dots, red blocks, yellow diamonds, purple pentagons, and blue hexagons are the results of the original images, this work, SBD, K-SVD, and LMC, respectively.}\label{compare}
\end{figure}
Other deep learning methods, such as cycle-consistent GAN (cycleGAN) \cite{zhu2017unpaired}, variational autoencoder (VAE) \cite{hou2017deep}, and U-shape convolutional network (U-Net) \cite{ronneberger2015u}, could also be potentially able to handle this low-bit-depth OCT reconstruction task, so we also investigate their performance using the exactly same training and testing configuration as those described in Section 2.3.\\
\indent Figure~\ref{comDeep} shows the representative OCT B-scans processed via these methods and the pix2pixGAN adopted in this work. From left to right are the original images, the reconstruction results using the cycleGAN, VAE, U-Net, and the pix2pixGAN adopted in this work, and the ground truth. Row 1, 3, and 5 are the results using 4-bit, 5-bit, and 6-bit images, respectively, row 2, 4, and 6 are their corresponding zoom-in views inside the red boxes. As demonstrated in the figure, the cycleGAN could reconstruct the OCT images well at the bit depths of 5 and 6, but work poorly at the bit depth of 4. The VAE, on the other hand, is only capable of recovering the global structure of retina but unable to reconstruct the tissue texture. Different from these two methods, the U-Net and the pix2pixGAN adopted in this work are capable of reconstructing the OCT images at each bit depth. However, the U-Net tends to create a denoising effect on the images and lose the detailed information of blood vessels and their projection shadows. \\
\indent We further calculated the quantitative metrics of the images processed by these methods as demonstrated in the left column of Fig.~\ref{compare}. From top to bottom are the PSNR, MSSSIM, and CORR2 as functions of the bit depth.  The black dots, red blocks, blue diamonds, purple pentagons, and yellow pentagons are the results of the original images, this work, U-Net, cycleGAN, and VAE, respectively. In accordance with the visual comparison in Fig.~\ref{comDeep}, the pix2pixGAN adopted in this work achieves the best performance among different bit depth. Even though its advantages decrease at higher bit depths because the original images are closer to the ground truth. On the other hand, the U-Net achieves best PSNR because of its denoising effect as mentioned before, which does not represent its superiority in this reconstruction task.
\subsection{Comparison with compressed sensing and sparse representation methods}
\begin{figure}[h!]
\centering\includegraphics[width=11cm]{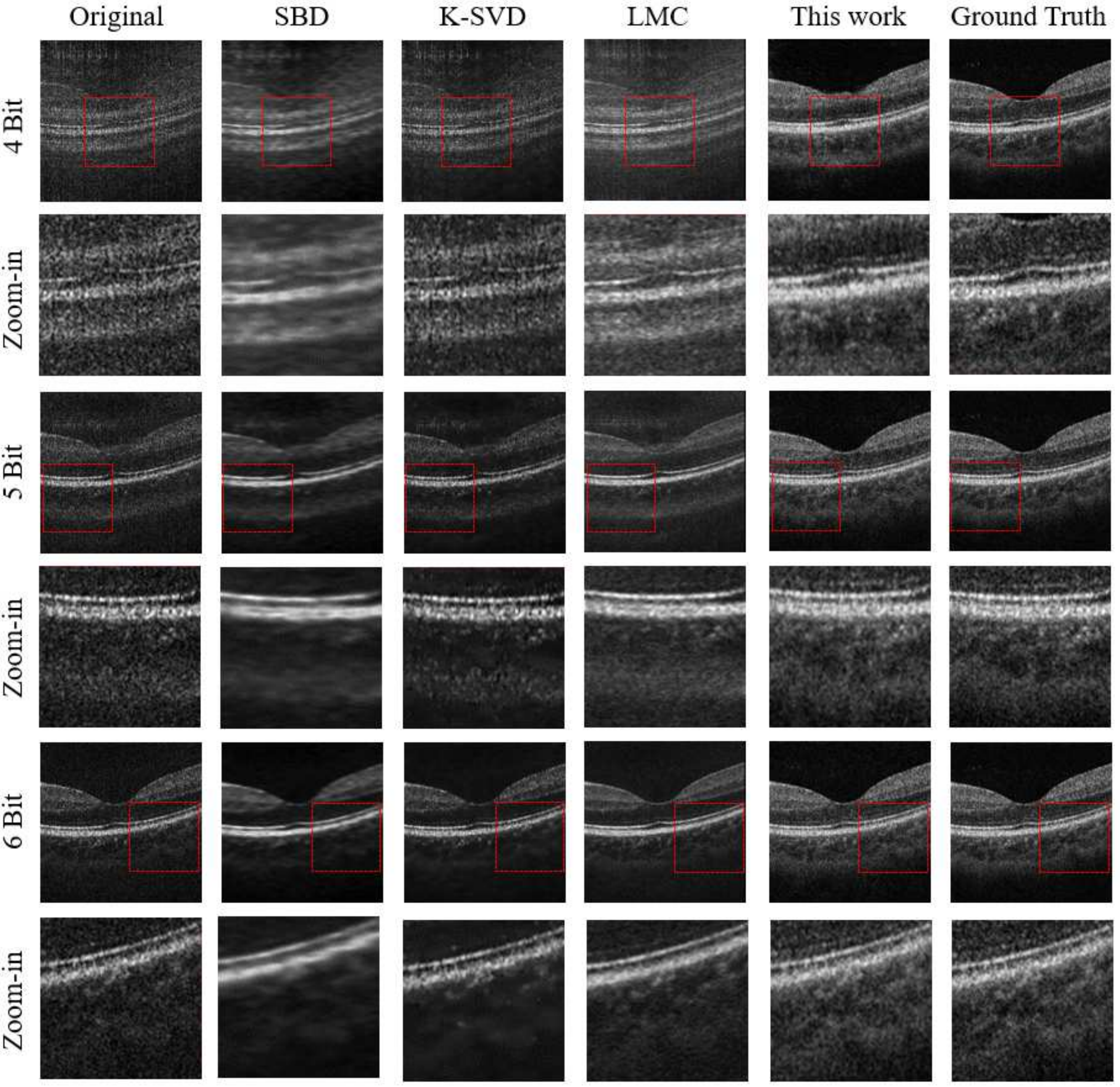}
\caption{Visual examples of the low-bit-depth OCT reconstruction using different sparse representation methods. From left to right: original images, the reconstruction results using the SBD, K-SVD, LMC, and the pix2pixGAN adopted in this work, and the ground truth. Row 1, 3, and 5 are the results using 4-bit, 5-bit, and 6-bit images, respectively, row 2, 4, and 6 are their corresponding zoom-in views inside the red boxes.}\label{comSparse}
\end{figure}
\begin{figure}[h!]
\centering\includegraphics[width=11cm]{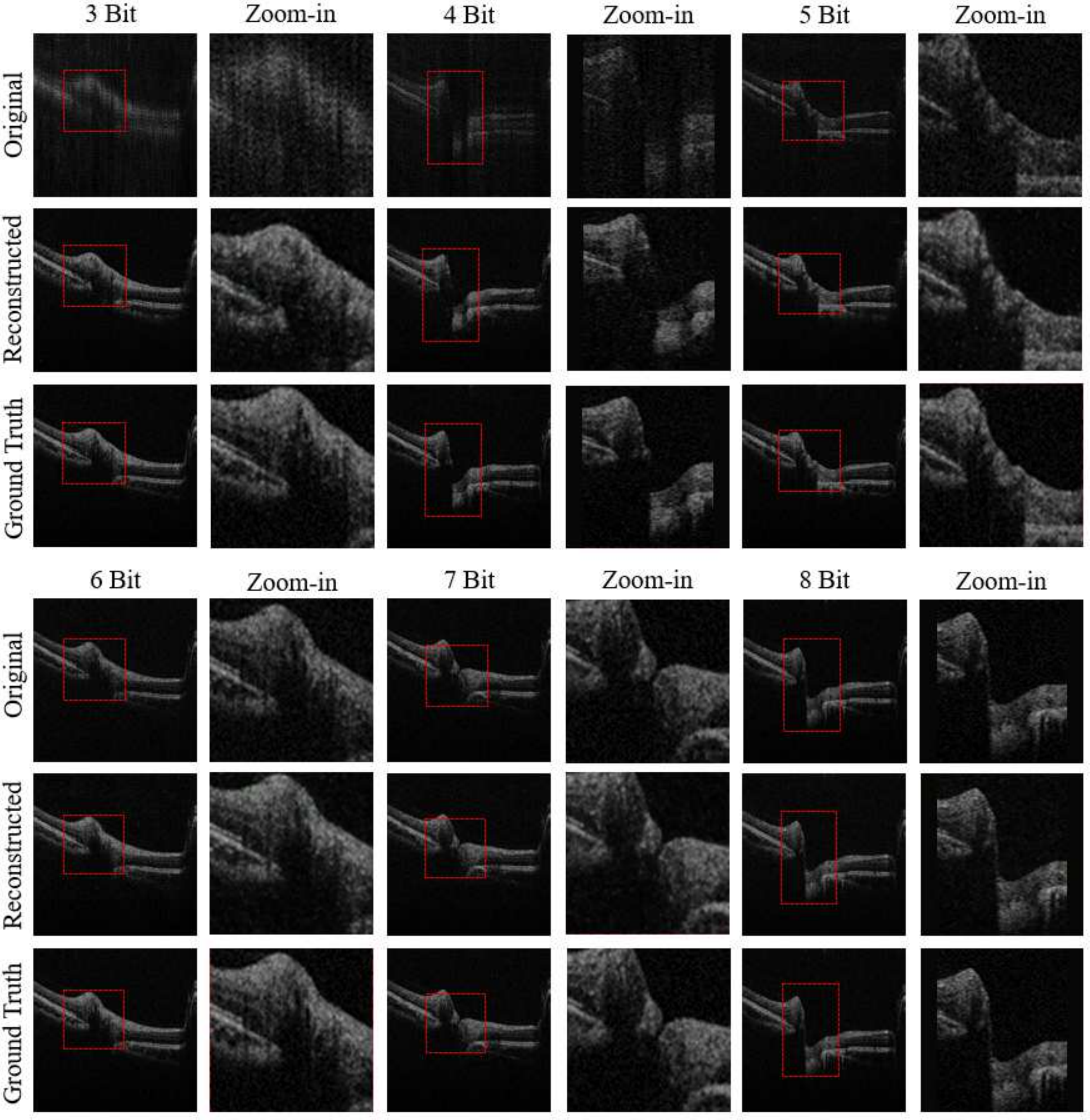}
\caption{OCT B-scans centered on the ONH at different bit depths and their corresponding reconstructed images and 12-bit ground truth. Column 2, 4, and 6 are the zoom-in views inside the red boxes.}\label{disc}
\end{figure}
\begin{figure}[h!]
\centering\includegraphics[width=11cm]{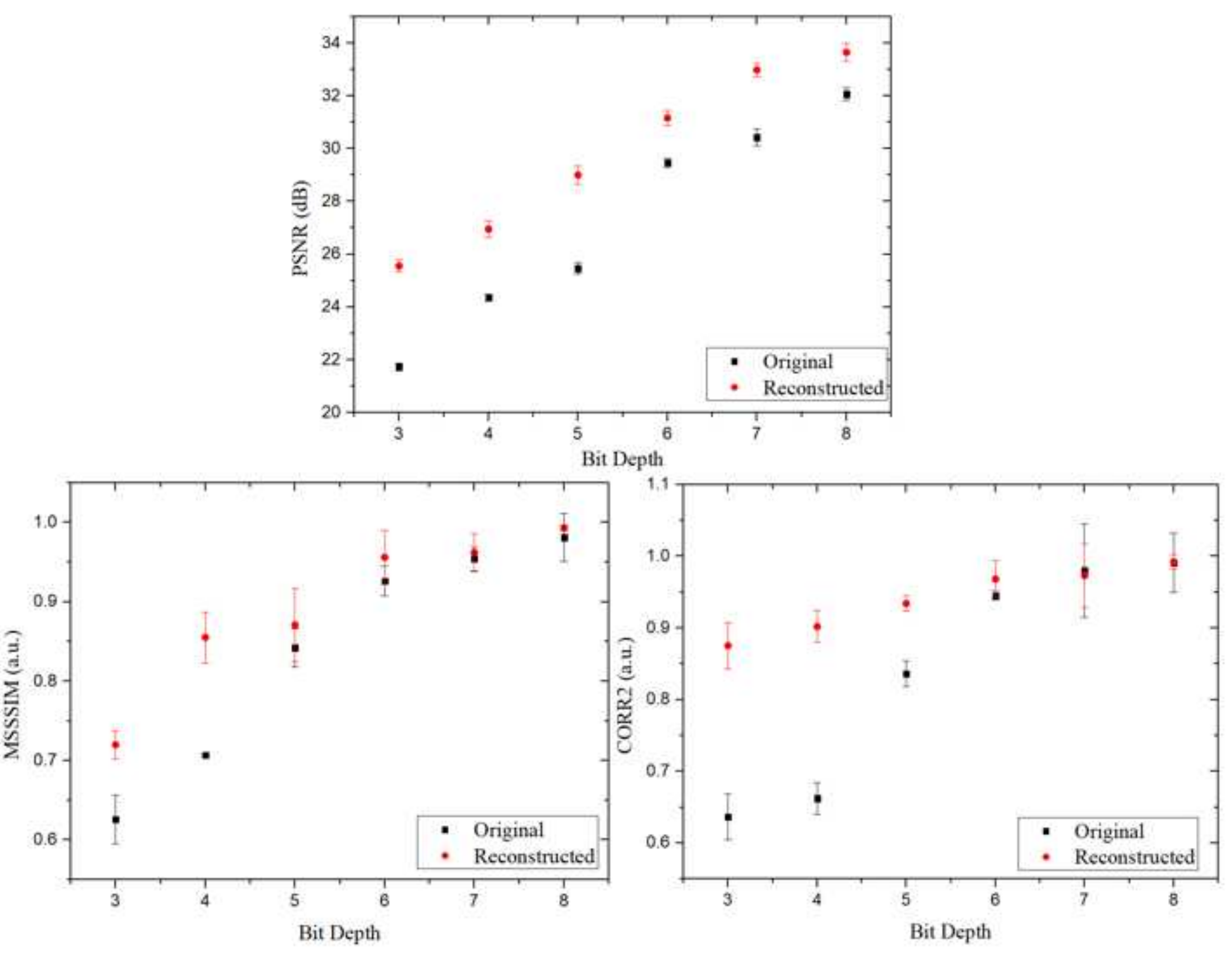}
\caption{Quantitative metrics for the original and reconstructed ONH images. The PSNR, MSSSIM and CORR2 as functions of bit depth. The black boxes represents the values of the original images. The red dots represents the values of the reconstructed images.}\label{discComp}
\end{figure}
Compressed sensing and sparse representation methods have also been used in the OCT reconstruction including low-spatial-sampling recovery and speckle noise reduction \cite{Liu2010,Lebed2010,Young2011,Zhang2012,Fang2013,cheng2016speckle,kafieh2014three}, so we also compare the pix2pixGAN adopted in this work with the these methods. Specifically, we employ the sparse-based denoising (SBD) \cite{Fang2013}, wavelet based singular value decomposition (K-SVD) \cite{kafieh2014three}, and low-rank matrix completion (LMC) \cite{cheng2016speckle} in this comparison. \\
\indent Figure~\ref{comSparse} shows the visual examples of the low-bit-depth OCT reconstruction using different sparse representation methods. From left to right: original images, the reconstruction results using the SBD, K-SVD, LMC, and the pix2pixGAN adopted in this work, and the ground truth. Row 1, 3, and 5 are the results using 4-bit, 5-bit, and 6-bit images, respectively, row 2, 4, and 6 are their corresponding zoom-in views inside the red boxes. As shown in the figure, the compressed sensing and sparse representation methods are unable to reconstruct the low-bit-depth OCT images properly. Only the enhancement of SNR could be observed in some cases. We could not conclude these methods are invalid for this task, because they may be effective in wave-number space, where the bit-depth undersampling happens. However, the further exploration of the OCT reconstruction in wave-number space is outside the scope of this work and makes the comparison of this image to image transition task unfair.\\
\indent The right column of Fig.~\ref{compare} are the results using compressed sensing and sparse representation methods. The black dots, red blocks, yellow diamonds, purple pentagons, and blue hexagons are the results of the original images, this work, SBD, K-SVD, and LMC, respectively. In accordance with the visual observation in Fig.~\ref{comSparse}, the compressed sensing and sparse representation methods have negative influence on this reconstruction task, while the pix2pixGAN adopted in this work has the best performance for all of the quantitative metrics.
\subsection{Generalization performance}
To verify the generalization performance of the proposed low-bit-depth OCT reconstruction using the pix2pixGAN method, we acquired 120 B-frames centered on the optic nerve head (ONH) of the same subject, which have significantly anatomical diversity compared with those scans centered on fovea (as shown in Fig.~\ref{disc}). The low-bit-depth ONH B-scans were generated using the same method described in Section 2.1. Then we retrained the model from scratch using a total of 320 B-scans. We randomly selected 250, 35, and 35 images for training, validation, and testing, respectively.\\
\indent Figure~\ref{disc} demonstrates several ONH B-scans at different bit depths and their corresponding reconstructed images and 12-bit ground truth. Column 2, 4, and 6 are the zoom-in views inside the red boxes. We can see the low-bit depth ONH images exhibit similar blurry and SNR degradation as those in the macular images. The reconstructed images, on the other hand, show excellent morphological and texture similarity with the ground truth, which suggest that this reconstruction approach is able to handle OCT images with significant morphological diversity.\\
\indent We further justify the visual observation by calculating the quantitative metrics. The PSNR, MSSSIM, and CORR2 as functions of the bit depth are shown in Fig.~\ref{discComp}. The black boxes represents the values of the original images. The red dots represents the values of the reconstructed images. In accordance with the trends in Fig.~\ref{fig6}, which is the quantitative metrics of the macular images, the GAN reconstruction significantly improves the image quality. As the bit depth increase, the difference between the original and reconstructed images decreases, because the quality of original images improves.
\subsection{Application in choroid segmentation}
\begin{figure}[h!]
\centering\includegraphics[width=10cm]{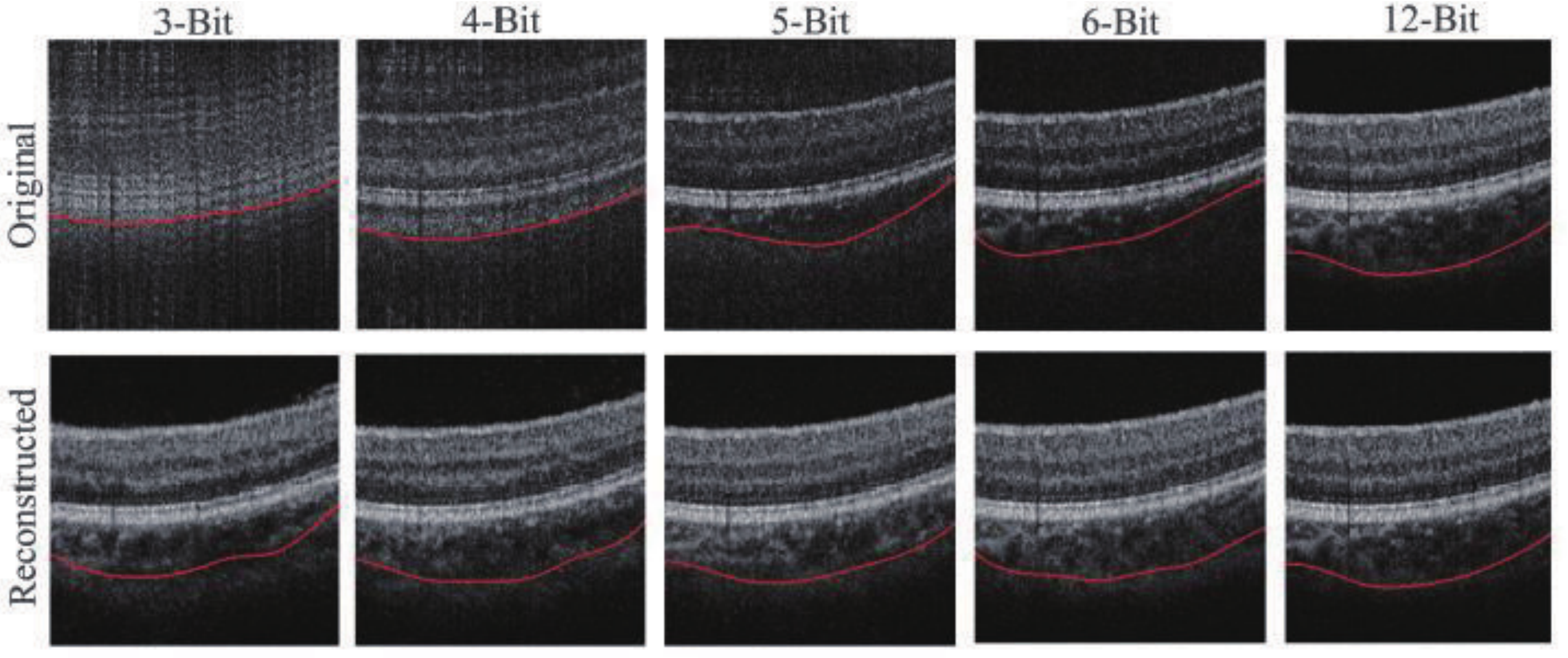}
\caption{The segmentation results of the CSI (red lines) using the original and reconstructed images with different bit depths.}\label{csi}
\end{figure}
\begin{figure}[h!]
\centering\includegraphics[width=11cm]{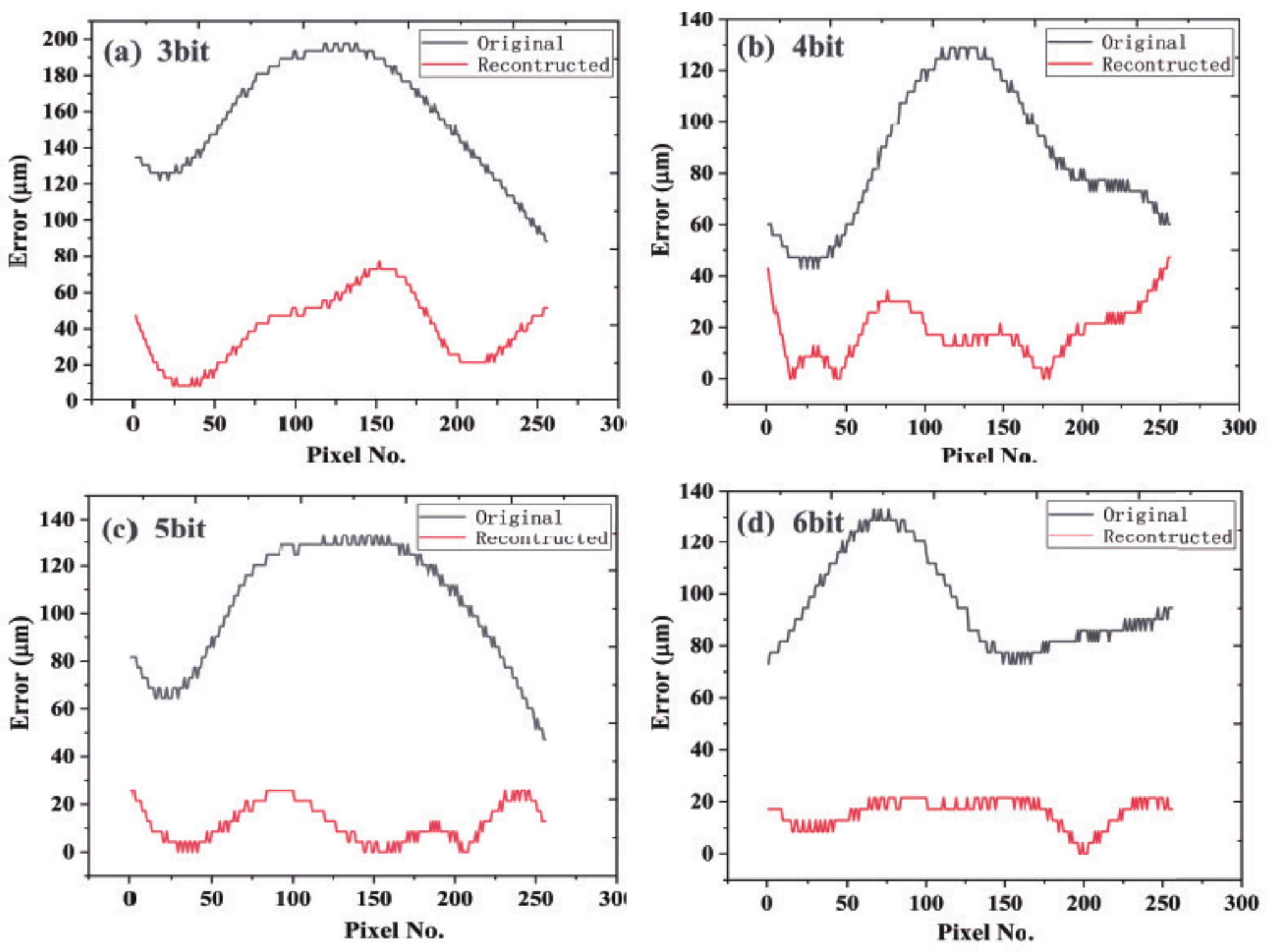}
\caption{The segmentation errors of the original (black) and reconstructed (red) images with the bit depths of $3\sim6$.}\label{seg}
\end{figure}
\indent To further prove the improvement of the SNR and similarity brought by this deep-learning-based reconstruction method, we employed a graph-search-based algorithm \cite{mazzaferri2017open} to segment the choroid-sclera interface (CSI) from the original and reconstructed OCT B-scan with different bit depths. The segmentation of the CSI is the most challenging task among the retinal layers because the OCT probe light is severely attenuated before reaching this layer. Also, the choroid is a vascular layer thus the boundary is composed of large vessels instead of the membranes separating other retinal layers. The light attenuation and the vascular boundary make the CSI very fuzzy. Besides, as mentioned above, the low bit-depth would cause the reduction of the SNR especially at the choroidal region. If the reconstruction succeeds, the SNR of the images will be improved, which further leads to the accurate segmentation of the CSI.\\
\indent Figure~\ref{csi} shows the representative segmentation results of the CSI using different bit-depth B-scans. The red lines indicate the positions of the segmented CSI. We can see the segmentation is very inaccurate in the original low bit-depth images because of the blur or low SNR. After the GAN-based reconstruction, the segmentation is significantly improved because the CSI can be clearly visualized in each image. As the bit-depth increases, the segmentation is closer to that of the 12-bit image.\\
\indent We set the automatically-segmented and manually-checked CSI of 12-bit B-scan as the ground truth. Using it as the reference, the segmentation errors of the original and reconstructed images were plotted in Fig.~\ref{seg}. For each bit depth, the errors are significantly decreased using the reconstructed image compared with the errors of the original image. For the reconstructed images, the average segmentation error decreases as the bit depth increases from 39.86 $\mu$m at 3 bit to 18.13 $\mu$m at 6 bit.
\section{Discussion and conclusions}
Using the low bit-depth acquisition could effectively lower the sampling and transmission requirements of OCT data, which will further contribute to reducing the cost of clinical OCT systems and the popularization of telemedicine. But it also has the side effect of degrading the image quality. The emerging of deep learning techniques bring the opportunities to reduce the bit-depth while keep the high SNR and resolution of OCT images. In this paper, we have preliminarily investigated the feasibility of this idea using an adversarial network to reconstruct the high SNR OCT images using their low bit-depth counterparts. Using the native 12-bit OCT images as the reference, we have found that the GAN-reconstructed images could achieve excellent structure and texture similarity especially when the bit-depth of the original images are $\geq 5$.  \\
\indent The success of this high SNR reconstruction of the low bit-depth OCT images suggests the implementation of the proposed idea (as demonstrated in Fig.~\ref{fig1}) can be safely moved to the next stage [Fig.~\ref{fig1}(b)]: the reconstruction from low bit-depth interference fringes to high SNR OCT images.  In this reconstruction from OCT raw photoelectric signals to B-scan images,  we need to train the DNNs to learn not only the features of different bit depths, but also the characteristics of the entire OCT post-processing including background subtraction, $k$-linearization, dispersion compensation, Fourier transformation, and image logarithm. So it may not be enough to use the current GAN architecture, we need to investigate the possibilities of infusing other deep learning architectures, CS techniques, and OCT knowledge for achieving high-quality reconstruction. The progressive realization of the proposed technique will benefit the development of healthcare in low-resource settings and telemedicine.\\
\indent In summary, We have proposed and implemented a deep learning based approach to reconstruct the low bit-depth OCT images for achieving high definition and SNR. Since the structure and texture information of the low bit-depth OCT B-scans and the native 12-bit OCT B-scans have precisely correspondence, we have adopted a pixel-to-pixel GAN architecture in the reconstruction. The GAN-generated images have achieved qualitatively and quantitatively accordance with the native 12-bit OCT images. We have further applied the reconstruction images in the segmentation of the CSI and achieved significant improvement in accuracy. 
\section*{Funding}
Ningbo ``2025 S\&T Megaprojects'' (2019B10033); Zhejiang Provincial Natural Science Foundation (LQ19H180001); Ningbo Public Welfare Science and Technology Project (2018C50049).

\section*{Acknowledgments}
We would like to thank the reviewers and editors for the careful reviewing and insightful comments, which significantly benefits the improvement of this manuscript.

\section*{Disclosures}
The authors declare no conflicts of interest.

\bibliography{sample}

\end{document}